\pdfoutput=1
\documentclass[fleqn,usenatbib]{mnras}

\usepackage{newtxtext,newtxmath}

\usepackage[T1]{fontenc}
\usepackage{ae,aecompl}

\usepackage{graphicx}	
\usepackage{amsmath}	
\usepackage{amssymb}	

\title[The curious activity in the nucleus of NGC 4151]{The curious activity in the nucleus of NGC\,4151: jet interaction causing variability?}

\author[D.R.A. Williams et al.]{D.R.A. Williams,$^{1,2}$\thanks{E-mail: david.williams@physics.ox.ac.uk}
R.D.Baldi,$^{2,4,5}$
I.M. McHardy,$^{2}$
R.J. Beswick,$^{3}$
F. Panessa,$^{5}$
\newauthor{D. May,$^{6}$
J. Mold{\'o}n,$^{3,7}$
M.K. Argo,$^{3,8}$
G. Bruni,$^{5}$,
B.T. Dullo,$^{9}$
J.H. Knapen,$^{10,11,12}$}
\newauthor{
E. Brinks,$^{13}$
D.M. Fenech,$^{14}$
C.G. Mundell,$^{15}$
T.W.B. Muxlow,$^{3}$
M. Pahari,$^{2}$
}
\newauthor{
J. Westcott$^{13}$
}
\\
$^{1}$Department of Physics, University of Oxford, Denys Wilkinson Building, Keble Road, Oxford, OX1 3RH, UK \\
$^{2}$School of Physics and Astronomy, University of Southampton, Southampton, SO17 1BJ, UK\\
$^{3}$Jodrell Bank Centre for Astrophysics, School of Physics and Astronomy, The University of Manchester, Manchester, M13 9PL, UK\\
$^{4}$Dipartimento di Fisica, Universit\'a degli Studi di Torino, via Pietro Giuria 1, 10125 Torino, Italy \\
$^{5}$INAF - Istituto di Astrofisica e Planetologia Spaziali, via Fosso del Cavaliere 100, I-00133 Roma, Italy \\
$^{6}$Instituto de Astronomia, Geof\'{i}sica e Ci\^{e}ncias Atmosf\'{e}ricas, Universidade de S\~{a}o Paulo, 05508-090 S\~{a}o Paulo, SP, Brazil\\
$^{7}$Instituto de Astrof\'isica de Andaluc\'ia (IAA, CSIC), Glorieta de las Astronom\'ia, s/n, E-18008 Granada, Spain\\
$^{8}$University of Central Lancashire, Jeremiah Horrocks Institute Preston, UK PR1 2HE\\
$^{9}$Departamento de F\'isica de la Tierra y Astrof\'isica, Instituto de F\'isica de Part\'iculas y del Cosmos IPARCOS, \\Universidad Complutense de Madrid, E-28040 Madrid, Spain\\
$^{10}$Instituto de Astrof\'{i}sica de Canarias, V\'{i}a L\'{a}ctea S/N, E-38205 La Laguna, Spain\\
$^{11}$Departamento de Astrof\'{i}sica, Universidad de La Laguna, E-38206 La Laguna, Spain\\
$^{12}$Astrophysics Research Institute, Liverpool John Moores University, IC2, Liverpool Science Park, 146 Brownlow Hill, \\Liverpool,
L3 5RF, UK\\
$^{13}$Centre for Astrophysics Research, University of Hertfordshire, College Lane, Hatfield, AL10 9AB, UK\\
$^{14}$Astrophysics Group, Cavendish Laboratory, 19 J. J. Thomson Avenue, Cambridge CB3 0HE, UK\\
$^{15}$Department of Physics, University of Bath, Claverton Down, Bath, BA2 7AY, UK\\
}

\date{Accepted XXX. Received YYY; in original form ZZZ}

\pubyear{2015}

\begin{document}
\label{firstpage}
\pagerange{\pageref{firstpage}--\pageref{lastpage}}
\maketitle

\begin{abstract}
A key characteristic of many active galactic nuclei (AGN) is their variability, but its origin is poorly understood, especially in the radio domain. Williams et al. (2017) reported a $\sim$50 \textit{per cent} increase in peak flux density of the AGN in the Seyfert galaxy NGC\,4151 at 1.5\,GHz with the e-MERLIN array. We present new high resolution e-MERLIN observations at 5\,GHz and compare these to archival MERLIN observations to investigate the reported variability. Our new observations allow us to probe the nuclear region at a factor three times higher-resolution than the previous e-MERLIN study. We separate the core component, C4, into three separate components: C4W, C4E and X. The AGN is thought to reside in component C4W, but this component has remained constant between epochs within uncertainties. However, we find that the Eastern-most component, C4E, has increased in peak flux density from 19.35$\pm$1.10 to 37.09$\pm$1.86 mJy/beam, representing a 8.2$\sigma$ increase on the MERLIN observations. We attribute this peak flux density increase to continued interaction between the jet and the emission line region (ELR), observed for the first time in a low-luminosity AGN such as NGC\,4151. We identify discrete resolved components at 5\,GHz along the jet axis, which we interpret as areas of jet-ELR interaction.

\end{abstract}




\section{Introduction}

Astrophysical radio jets are a common signature of the accretion process onto compact objects such as super-massive black holes (SMBHs). Jets are responsible for mechanical ``feedback'' into their surroundings, triggering and quenching star formation as well as regulating galaxy evolution \citep[e.g. ][]{Fabian2012,Morganti2013}. An accreting SMBH is known as an active galactic nucleus (AGN).
The most powerful AGN such as those in Fanaroff-Riley Type I and Type II radio galaxies \citep{fanaroff74} host jets that can be launched at close to the speed of light \citep{begelman84,UrryPadovani95}. However, in low-luminosity AGN (LLAGN), and radio-quiet AGN 
\citep[defined by ][as L$\rm _{radio}$/L$\rm _{opt}$ $<$ 10]{Kellerman1989}, compact radio emission and jet-like features have been observed, though often less collimated and less powerful \citep[e.g.][]{nagar00,BaldiLeMMINGs}. In such sources, the presence of compact nuclear radio emission has been attributed to a variety of different processes \citep[see][for a review]{Panessa2019}, including free-free emission/absorption \citep{UlvestadHo2001,Gallimore2004}, a combination of a compact radio jet plus an advection dominated accretion flow \citep[ADAF, ][]{FalckeMarkoff2000}, a standard geometrically-thin optically-thick accretion disk \citep{ShakuraSunyaev,Ghisellini2004} or outflowing material at relatively low velocities \citep{Giroletti2017,Laor2019}. To resolve these different emission processes in nearby LLAGN, high-resolution radio interferometers such as e-MERLIN are necessary to separate the core component from circum-nuclear star formation.

Radio variability has been observed in luminous AGN \citep[e.g.][]{Hufnagel1992} and is a useful tool for discriminating between the different radio emission mechanisms \citep{Panessa2019}. In some cases, the radio variability is found to be related to activity in the jet, rather than in the self-absorbed synchrotron emitting core component attributed to the AGN \citep[e.g. in NGC\,1052, ][]{Baczko2016,Baczko2019}. However, due to their intrinsic radio-weakness, variability studies of LLAGN have been limited to small samples of the best known sources, predominantly Seyfert galaxies \citep{wrobel00,Nagar2002,Mundell2009,Bell2011}. Variability of up to 40 \textit{per cent} has been reported in half of a sample of Seyfert galaxies over a seven year period \citep{Mundell2009}, while almost no variability has been reported in some other cases \citep[e.g. NGC\,4051,][]{SadieJones2011}. Shorter, month-long variability has been detected in the Seyfert galaxy NGC\,5548, \citep{wrobel00}, and weak radio variability has also been detected on a few days timescales in radio-quiet AGN systems \citep{anderson05,Baldi2015}. However, only a handful variability studies have been undertaken at decade-long timescales or longer and in most cases the variability has been attributed to the emission from the core.

\begin{figure*}
\centering
\includegraphics[width=0.99\textwidth]{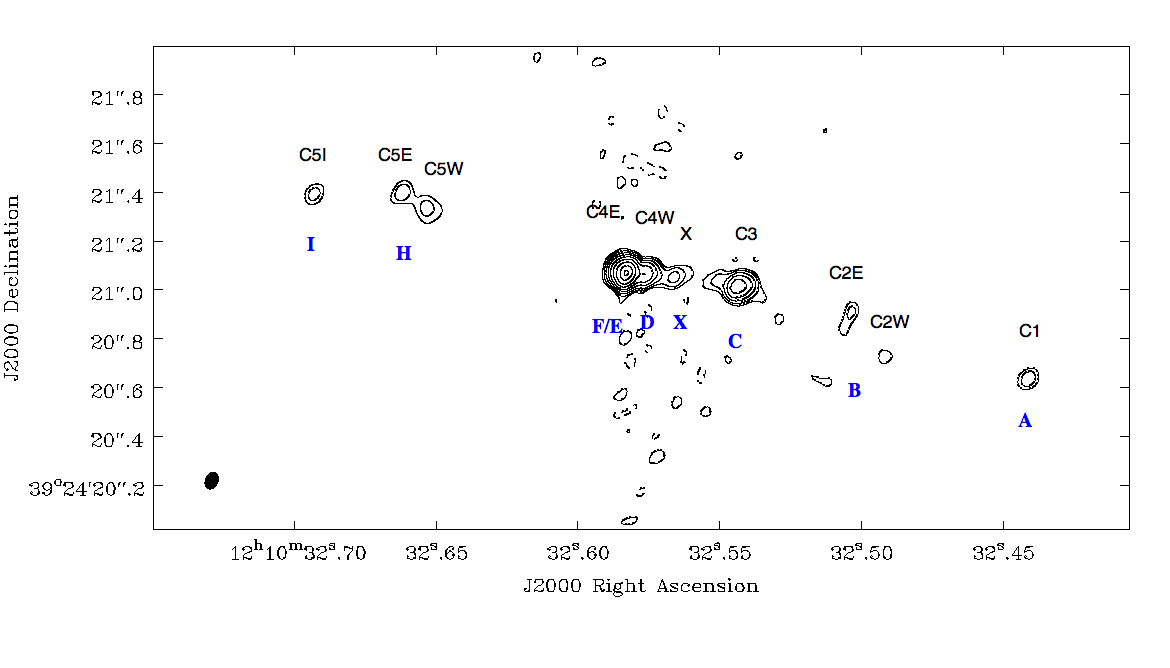}
\caption{Naturally weighted full resolution e-MERLIN 5\,GHz image of the central 2.0$\arcsec \times$3.75$\arcsec$ ($\sim$180$\times$340\,pc) of NGC\,4151. The e-MERLIN synthesized beam is shown as the filled ellipse in the bottom left hand corner: 0.07$\arcsec \times$0.05$\arcsec$ with a P.A. of -24$^{\circ}$. Contours are set at the rms noise level, 70$\mu$Jy/beam $\times$ -2.5, 2.5, 5, 10, 20, 40, 80, 160, 320, 500. 
The bold, blue, labels below (i.e., south of) the components (A to I) refer to the labels from higher resolution VLBA observations \protect{\citep{Mundell03}}. The black labels above (i.e., north of) the components (C1 to C5) refer to components from lower resolution MERLIN and e-MERLIN observations \protect{\citep{Pedlar93,Williams4151}}. The component properties are listed in Table 1.}
\label{fig:BigMap}
\end{figure*}

Recently, a change of radio flux density has been reported for the core of the radio-quiet Seyfert galaxy NGC\,4151. In MERLIN and e-MERLIN resolution maps \citep{Pedlar93,Williams4151} the components are labelled from C1 to C5 and the unresolved core is C4. Higher resolution VLBI observations \citep{Ulvestad98,Mundell03,Ulvestad2005} resolve some of these components into multiple smaller components. The VLBI components are labelled A to I. The correspondence between the VLBI and MERLIN/e-MERLIN components is shown in Fig.1 of \citet{Mundell03}. In this paper we use both conventions, which we also illustrate in Fig.\ref{fig:BigMap} here. The central component, C4, unresolved in e-MERLIN observations at 150 mas at 1.5 GHz, increased in peak flux density from $\sim$37\,mJy/beam \citep{Mundell95} to $\sim$67\,mJy/beam \citep{Williams4151}, over a period of 22 years. 
We obtained new 5\,GHz observations of NGC\,4151 with e-MERLIN (proposal ID: CY6219), which probes angular resolutions up to 50\,mas, improving the angular resolution of the observations of \cite{Williams4151} by a factor of three. This allows us to investigate the varying radio components in the galaxy core at higher imaging fidelity.

%

We assume a distance to NGC\,4151 of 19\,Mpc, \citep{Honig14}. This paper is structured as follows: In Section 2 we describe the new e-MERLIN data and reduction procedures. In Section 3 we show our results and in Section 4, discuss their implications on the AGN core position in NGC\,4151 and Section 5 lists our conclusions.

\section{Observations and Data Reduction}
\label{sec:Obs}

NGC\,4151 was observed on 2017 July 19 and correlated at Jodrell Bank, using all e-MERLIN stations except the Lovell telescope. The observing set-up was centred at 5.07\,GHz, using a total bandwidth of 512\,MHz and split into four spectral windows (spws). The target field was observed for 3.15 hours in 6.5 minute scans, interleaved with 2.5 minute scans of the phase calibrator NVSSJ115354+403652 (J1153+4036, RA: 11$^{\rm h}$53$^{\rm m}$54.6594$^{\rm s}$, Dec: +40$^{\circ}$36$\arcmin$52.617$\arcsec$). The standard e-MERLIN band pass (OQ208) and flux density (3C286) calibrators were observed at the beginning of the observing run for 14 minutes and 25 minutes respectively. Unfortunately, the data for the Mark II telescope did not provide any phase information and thus the antenna was flagged; Pickmere was used as a reference antenna. The remaining five antennas generally showed good phase stability. The data were calibrated with version 0.7.9 of the \textsc{CASA} e-MERLIN pipeline\footnote{Online documentation can be found at: https://github.com/e$-$merlin/CASA$\_$e$-$MERLIN$\_$pipeline/wiki}. The \textsc{CASA} e-MERLIN pipeline is similar to the previous \textsc{AIPS} data reduction procedure outlined in \citet{Williams4151}, \citep[see also][]{Westcott2017,BaldiLeMMINGs,Williams2019} for e-MERLIN continuum data: the pipeline performs standard calibration of e-MERLIN data by flagging radio frequency interference (RFI) using \textsc{AOFlagger} \citep{AOFlagger}, solving the delays of the antennas, performing phase and amplitude calibration, bootstrapping the flux density scaling, fitting and applying a band pass and making preliminary images and diagnostic plots, ready for further imaging and self calibration of the target fields.

The imaging procedures were conducted in \textsc{CASA} using the \textsc{CASA} task \textsc{tclean}. We performed a standard phase self-calibration on the target source and found some amplitude errors on the shortest baselines between the Pickmere and Darnhall telescopes, which we then flagged from the data. By removing this information, we are less sensitive to the diffuse emission along the jet, but are still able to resolve small compact components of emission. When no further improvements through self-calibration could be made, a final image was produced with natural weighting. An rms noise level of 70$\mu$Jy/beam was achieved for the e-MERLIN data and was estimated from a region near the phase centre but not including the radio components of the jet. The rms noise regions were chosen so as not to encompass the areas directly north and south of the core region, which show some low-level artefacts in the image due to lack of \textit{uv}-coverage on the shortest baselines. The full resolution image, with a synthesized beam of 0.05$\arcsec \times$0.07$\arcsec$and a position angle (P.A.) of -24$^{\circ}$, is shown in Fig.~\ref{fig:BigMap}. The P.A. of the full resolution image resulted in the major axis of the beam being perpendicular to the jet axis, allowing for identification of jet structures down to 0.05$\arcsec$, corresponding to a linear scale of $\sim$4.6\,pc. 

\subsection{Archival MERLIN data}
NGC\,4151 was observed by MERLIN, the predecessor to e-MERLIN, on 1991 September 12 at 4.993\,GHz with a bandwidth of 7\,MHz. These data are published in \citet{Pedlar93} at a reduced resolution of 75\,mas. We refer the reader to \cite{Pedlar93} for the details of the 1991 observation and data reduction procedures. We obtained the calibrated Pedlar et al. archival data from Jodrell Bank with the goal of imaging them at the same resolution and \textit{uv}-range as our new e-MERLIN data, to ensure that any changes of observed flux density found with the new data were significant (see next Section). The calibrated \textit{uv}-dataset was examined and self calibrated in \textsc{AIPS} and exported into \textsc{CASA} file format for commensurate cleaning procedures with the e-MERLIN data. 

\subsection{\textit{uv}-matching and imaging the datasets}

To ensure that any changes in flux density are genuine and comparison between the two epochs is robust, we ensured that the e-MERLIN and MERLIN datasets used the same \textit{uv}-information \citep[see Section 2.1.3 of][for details of the procedure performed for the 1.5\,GHz data]{Williams4151}. As the \textit{uv}-coverage of the new e-MERLIN array is better than that of MERLIN, we first matched the \textit{uv}-coverage of the two datasets, using the same \textit{uv}-range (400-3500k$\lambda$). This range was chosen to use all of the long baseline information to ensure the highest possible resolution images, but to remove amplitude errors found on the shortest baseline between Pickmere and Darnhall in the e-MERLIN data. As a consequence, this \textit{uv}-range corresponds to angular scales of 0.05-0.5$\arcsec$, which means the new e-MERLIN data are not sensitive to diffuse emission larger that 0.5$\arcsec$ in size. We removed antennas from both datasets that only participated in one observation\footnote{The now defunct Mark III (Wardle) antenna did not take part in the 1991 observations, and we removed the Mark II antenna from the 1991 data as it was not present in the 2017 e-MERLIN data.}. Finally, we fixed the restoring beam of both datasets to the same major and minor axes, as well as position angle. As the full resolution MERLIN data resulted in a circular beam of 0.05$\arcsec$, we cleaned both datasets with a restoring beam of 0.07$\arcsec$ $\times$ 0.05$\arcsec$ and a P.A. equal to that of the full resolution e-MERLIN data, in order to preserve as much resolution along the jet axis as possible, while avoiding a super-resolution of either dataset.

\begin{figure}
\centering

   \includegraphics[width=\columnwidth]{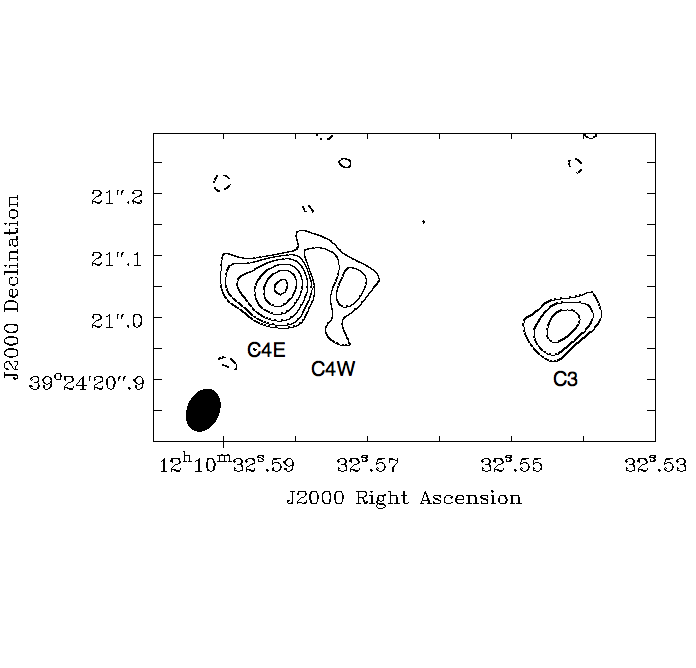}
   \caption{Archival MERLIN data from \protect{\citep{Pedlar93}} of the central 0.50$\arcsec \times$ 0.81$\arcsec$ of NGC\,4151, corresponding to a linear scale of 45$\times$74\,pc$^2$, re-imaged with a restoring beam of 0.05$\arcsec \times$0.07$\arcsec$ and a P.A. of -24$^{\circ}$ to match the e-MERLIN data, indicated by the filled ellipse in the corner of the image. The contour levels are set as the rms noise level, 0.5\,mJy/beam, $\times$ -3, 3, 5, 9, 16, 25, 36, 49, 64. }
   \label{fig:MERLIN} 

   \includegraphics[width=\columnwidth]{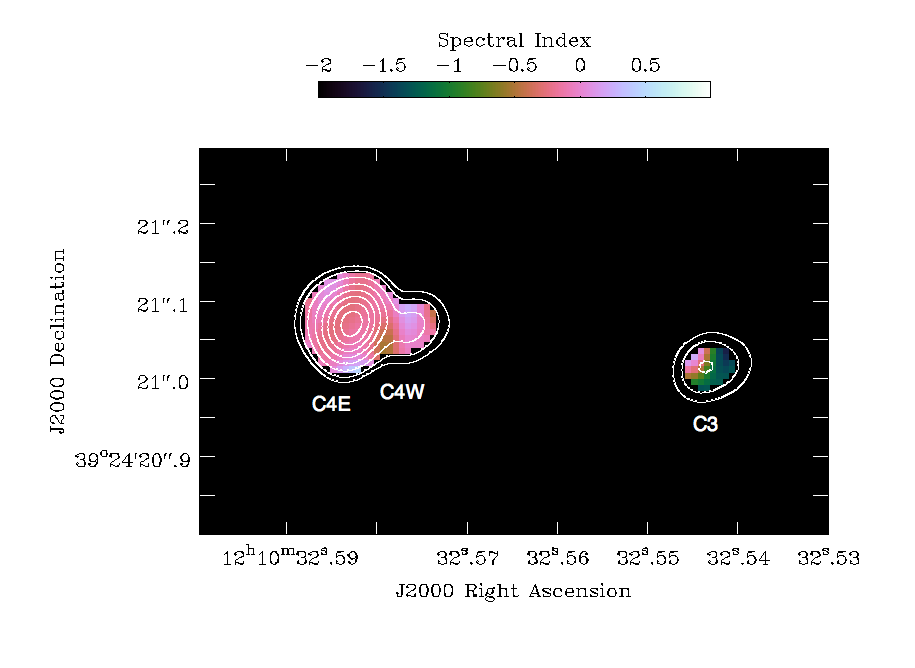}
   \caption{As in Fig.~\ref{fig:MERLIN}, but now presenting the new e-MERLIN 5\,GHz data of NGC\,4151. The contour levels, beam size and P.A. are the same as in Fig.~\ref{fig:MERLIN}. The greyscale intensity image shows the spectral index image described in Section~\ref{sec:spec}, with the intensity bar indicating the spectral index at the top. }
   \label{fig:eMERLIN}
\end{figure}

The final MERLIN image reached an rms-noise value of 500\,$\mu$Jy/beam. This noise is higher than that reported in \cite{Pedlar93} ($\sim$100$\mu$Jy/beam) and is likely due to the reduced \textit{uv}-range used in these data and removal of the Mark II telescope. As such, we are unable to make an image of the entire jet structure as the \textit{uv}-range cut and lack of short spacing antennas means we are no longer sensitive to the diffuse emission along the jet axis shown by \citet{Pedlar93}. We show the MERLIN image of the detected components C3 and C4 in Fig.~\ref{fig:MERLIN}.
We show the \textit{uv}-limited e-MERLIN image for the same region of the jet in Fig.~\ref{fig:eMERLIN}.

\subsection{Fitting components in the data}

The flux densities of all the components were extracted using the \textsc{IMFIT} task in \textsc{CASA}, which fits a two dimensional Gaussian to the regions of interest. For sources where the component profiles are merged or overlap, e.g. components C4E/C4W, we fix the source size to that of the beam and fit both the positions and flux densities of the sources simultaneously. The flux density uncertainties obtained from the fitting procedure are added in quadrature to a 5 \textit{per cent} flux density bootstrapping uncertainty for MERLIN/e-MERLIN data \citep{RampadarathM51}. We report the extracted flux densities and associated uncertainties from the \textit{uv}-matched datasets and are reported in Table~\ref{table:Table1}. 

\subsection{Radio Spectra}
\label{sec:spec}
As an additional analysis, we split the e-MERLIN band into two halves, in order to create an in-band spectral index image using the \textsc{CASA} task \textsc{immath} (grey-scale plot in Fig.~\ref{fig:eMERLIN}). The spectral index is defined as S$_{\nu} \propto \nu^{\alpha}$. As the e-MERLIN band at 5\,GHz is 512\,MHz, there is a decrease in sensitivity across the band of $\sim$7 \textit{per cent}, assuming a spectral index of -0.7 is used, applicable for radio galaxies. To ensure that the spectral index values are robust, we require that the difference across the band is at least 3 $\times$ the image rms noise level of 0.07\,mJy/beam. Therefore, as 7 \textit{per cent} of the band is at least 0.21\,mJy/beam, then the total flux density must be at least $\sim$3\,mJy/beam. Hence, we removed all spectral index values for pixels with a flux density below this level\footnote{see the e-MERLIN webpage for a step-by-step discussion of how to perform this calculation: $http://www.e-merlin.ac.uk/data_red/CASA/Errors.html$}. 
This step allows us to have confidence that the spectral index values for the detected components are robust and correct. The 3\,mJy/beam cut off only leaves in-band spectral index values for components C3, C4E and C4W. We estimate uncertainties on the spectral index of each component by fitting a Gaussian to the pixel values in the spectral index image and report the full-width at half maximum value as the 1$\sigma$ error. The spectral indices of the components detected above the 3$\sigma$ level are shown in Table~\ref{table:Table1}.

\section{Results}
\label{sec:Results}

We present the full resolution e-MERLIN image in Fig.~\ref{fig:BigMap}. Listed above the components are the component names used in this paper, which are based off the names from previous studies at lower resolution \citep{Carral90,Williams4151}. We add additional letters to distinguish Eastern and Western components. Below the components in Fig.~\ref{fig:BigMap}, we show the component names used in VLBA studies at higher resolution \citep{Ulvestad98, Mundell03}. The increased sensitivity of the new e-MERLIN observations over the archival MERLIN data allows us to investigate additional compact components along the jet axis (see Section~4.2). We now break the radio jets into three parts, to describe their morphological properties, flux densities and spectra. The Western jet contains components C1 through C2, the nuclear region includes components C3 and C4, and the Eastern jet as components C5 to C6.

\subsection{The Western Jet}
Component A (also known as C1), is the first component of the Western jet, furthest away from the nucleus. 
It has a peak flux density of 0.66$\pm$0.08\,mJy/beam, and a similar integrated flux density, indicating that it is unresolved. Component B splits into two separate compact components in our 5\,GHz data, which we label C2W and C2E. A hint of this break can be seen in the 1.5\,GHz e-MERLIN images (Fig.~3 in \citealt{Williams4151}), but is strengthened by the appearance of two regions in the VLBA/VLA images at 21\,cm. The Western component (C2W) is unresolved 
and is only detected at $\sim$4$\sigma$ at a peak flux density of 0.29$\pm$0.07\,mJy/beam in the e-MERLIN data. 
The Eastern component of C2 on the other hand is extended north-south, and has a peak flux density of 0.47$\pm$0.07\,mJy/beam. Due to the low flux density of all components in the Western jet, we were unable to ascertain a spectral index for these components.

\subsection{The Nuclear Region}

The \textit{uv}-matched MERLIN and e-MERLIN images of the central core components C3 and C4 are presented in Fig.~\ref{fig:MERLIN} and Fig.~\ref{fig:eMERLIN}, respectively. The brightest features of both these components are detected in both the archival MERLIN and the new e-MERLIN 5\,GHz data. C3 (C) is slightly resolved in the east-west direction and it has a peak flux density of 4.56$\pm$0.26\,mJy/beam in the new e-MERLIN data, which represents a 2.7$\sigma$ decrease from the archival MERLIN data, when it was 6.32$\pm$0.61\,mJy/beam. The in-band e-MERLIN spectral index is -1.19$\pm$0.76, indicating it is likely spectrally steep, but the uncertainties are large.

Component C4 is separated into four distinct components in the 21\,cm VLBA images: X, D (C4W) and E (C4E) and F. We detect all bar component F, which is likely connected with component E in the VLBA data (see Fig.~2 of \cite{Mundell03}), but is faint, diffuse and possibly also resolved out in our e-MERLIN data. Moving west-to-east, component X is faint but detected at $\ga$10$\sigma$, with a peak flux density of 0.92$\pm$0.08\,mJy/beam. It is slightly extended in the east-west direction. It should be noted that the full resolution images shown in \citealt{Pedlar93} hint at an extension that could plausibly be related to component X, but we do not detect this component in the re-reduced MERLIN images above a 3$\sigma$ threshold. This is likely due to the reduction in \textit{uv}-range for our images. As such, it is not possible to investigate whether it has varied between epochs. Similarly, this source is below the cut-off level in the spectral index e-MERLIN image, so we cannot draw any robust conclusions on the spectra in this component.

The AGN is thought to reside in Component C4W (D) (see Discussion), and it is detected in the MERLIN data (3.82$\pm$0.54\,mJy/beam) and the e-MERLIN data (5.38$\pm$0.28\,mJy/beam), indicating a tentative increase of 2.6$\sigma$ in flux density between epochs. This component also has a spectral index of -0.02$\pm$0.34, indicating it is spectrally flat. Component C4E is the brightest of the three detected components at a peak flux density of 37.09$\pm$1.86\,mJy/beam. In the MERLIN data, this component was detected at a peak flux density of 19.35$\pm$1.10\,mJy/beam, which indicates that this component has nearly doubled in the 25 year period between the two epochs and represents a change of 8.2$\sigma$. The radio spectrum of this source (-0.19$\pm$0.17) indicates that this component is possibly spectrally steeper than component C4W.

\subsection{The Eastern Jet}
C5, the largest component from \cite{Mundell03} is split into three components spanning $\sim$1$\arcsec$: G, H and I in the VLBA/VLA images. Component G is not detected in our e-MERLIN images to a 3$\times$rms-noise limit of 0.21mJy/beam, but it is one of the faintest and more diffuse components, which our data is less sensitive to. Component H separates into two components in our e-MERLIN images, which we delineate once more into east and west components. The Western component of H is detected at 0.49$\pm$0.08\,mJy/beam and the Eastern component has a peak flux density of 0.67$\pm$0.08\,mJy/beam. Finally, component I is detected also as a compact, unresolved point source with a peak flux density of 0.55$\pm$0.08\,mJy/beam. None of the components in the Eastern jet were detected in the re-analysed MERLIN data, and similar to the Western jet, and the flux densities in the e-MERLIN data are too low to reach a reliable conclusion for their spectra either.

\begin{table*}
\begin{tabular}{l l cl cl cl cl cl cl cl cl cl}
\hline
& e-MERLIN & e-MERLIN & e-MERLIN & e-MERLIN & MERLIN & MERLIN & Spectral & Peak Flux\\
Comp. & Right & Declination & Peak Flux & Int Flux & Peak Flux & Int Flux & Index & Density\\
& Ascension&  & Density  & mJy & Density & mJy & $\alpha$ & Change \\
&  &  &  mJy/beam &  & mJy/beam &  &  & ($\sigma$) \\
\hline
C1 (A) & 12 10 32.441 & +39 24 20.64 & 0.66$\pm$0.08 & 0.70$\pm$0.15 & - & - & - & -\\
C2W (B) & 12 10 32.491 & +39 24 20.73 & 0.29$\pm$0.07 & 0.34$\pm$0.14  & - & - & - & -\\
C2E (B) & 12 10 32.503 & +39 24 20.90 & 0.47$\pm$0.07 & 0.53$\pm$0.13  & - & - & - & -\\
\hline
C3 (C) & 12 10 32.543 & +39 24 21.02 & 4.56$\pm$0.26 & 7.31$\pm$0.49 & 6.32$\pm$0.61 & 9.50$\pm$1.29 & -1.19$\pm$0.76  & 2.7\\
C4 (X) & 12 10 32.566 & +39 24 21.05 & 0.92$\pm$0.08 & 1.36$\pm$0.20  & - & - &  -  & -\\
C4W (D) & 12 10 32.576 & +39 24 21.07 & 5.38$\pm$0.28 & 5.62$\pm$0.31  & 3.82$\pm$0.54 & 5.0$\pm$1.1 & -0.02$\pm$0.34  & 2.6\\
C4E (E) & 12 10 23.583 & +39 24 21.07 & 37.09$\pm$1.86 & 43.70$\pm$2.19  & 19.35$\pm$1.10 & 26.4$\pm$1.7 & -0.19$\pm$0.17  & 8.2\\
\hline
C5W (H) & 12 10 32.653 & +39 24 21.33 & 0.49$\pm$0.08 & 1.10$\pm$0.25  & - & - &   & -\\
C5E (H) & 12 10 32.662 & +39 24 21.40 & 0.67$\pm$0.08 & 0.86$\pm$0.16  & - & - &   & -\\
C5 (I) & 12 10 32.693 & +39 24 21.39 & 0.55$\pm$0.08 & 0.60$\pm$0.13  & - & - &   & -\\

\end{tabular}
\caption{New e-MERLIN 5\,GHz positions, flux densities and spectral indices of all components identified in Fig.~\ref{fig:eMERLIN}, as well as MERLIN flux densities for the detected components obtained from Fig.~\ref{fig:MERLIN}, all obtained from the \textit{uv}-matched datasets. The uncertainties on the flux densities are found from the 2D Gaussian fitting procedure in \textsc{CASA} called \textsc{IMFIT}. We report the flux density uncertainties which include the flux density boot strapping uncertainty from the flux density calibrator 3C286, thought to be no more than 5 \textit{per cent} in the MERLIN and e-MERLIN data, and the fitting uncertainty from \textsc{IMFIT}, added in quadrature. The flux density change is calculated using the equations in \citet{Zhou2006, Bruni2012}. The positions listed in this table are typically accurate to a $\sim$10\,mas, or approximately 20 \textit{per cent} of the synthesized beam.}

\label{table:Table1}
\end{table*}

\section{Discussion}

Previous VLBI studies of NGC\,4151 at GHz frequencies have suggested that the AGN resides unresolved inside either component C4E \citep{Pedlar93,Ulvestad98} or C4W \citep{Mundell95,Ulvestad05}. The first VLBI observations of NGC\,4151 showed that component E, unresolved inside C4E, was the brightest component at 1.5\,GHz and separates into several smaller components that are perpendicular to the overall jet axis \citep{Ulvestad98}. Hence it was thought that the jet emerged from the AGN in component E and then bent at nearly right angles to form the rest of the observed jet. However, further VLBI studies at multiple frequencies showed that the radio spectral index of component E was steep and therefore unlikely to be the site of the AGN \citep{Ulvestad05}. In the same work, component D, unresolved inside component C4W, was shown to separate into three distinct components. The central component of D3, D3b, has a flat spectral index, a flux density of 3.0$\pm$0.4\,mJy and a brightness temperature, $\rm T_b$ $>$ 10$^{8}$K, indicative of an AGN. Furthermore, in HI studies of NGC\,4151, \cite{Mundell95} showed that the absorbing HI column towards component E/C4E was several orders of magnitude larger than that towards component D/C4W, indicating that component E/C4E lay behind the obscuring medium of the torus and therefore could not be the AGN. They observed "banana-like" structures in components C4E and C3, indicative of interaction with dense clouds of gas in the emission-line region (ELR) close to the AGN. Thus the consensus now is that the AGN most likely resides in component C4W and that component C4E is the site of the first jet component that is interacting with the ELR.

Our new e-MERLIN data show that the radio component C4W has not significantly varied over the course of 25 years (2.6$\sigma$ change). Furthermore, it has a radio spectral index consistent with being an AGN, in agreement with the previous studies. The e-MERLIN data presented here are unable to challenge the interpretation of the AGN being in component C4W. 

\subsection{Explaining the variability in C4E}

Component C4E has clearly doubled in flux density and is responsible for the majority of the flux density increase observed in the lower-resolution e-MERLIN studies. In addition C4E has a slightly steeper radio spectral index, but is also consistent with being spectrally flat, given the uncertainties. We now consider what may be causing the change in flux density in C4E, focussing on processes not related to accretion. These include shock radio emission from interaction of the jet with dense clouds in the extended emission line region \citep[ELR, e.g.][]{Middelberg2007}, or continued acceleration of particles along the jet \citep[e.g.][]{Blandford2019}. If the flux density increase in C4E is interpreted as due to shocks from jet-ELR interaction, multi-frequency VLBI observations should reveal a significantly brighter component, an extended morphology to that observed in \cite{Ulvestad2005}, and possibly a flat spectrum `hot spot' of emission with a steeper surrounding spectral index. If the observed flux density change is due to further particle acceleration in the jet, we would expect a flat spectrum source. We note that our e-MERLIN data is not of high enough resolution to distinguish between these two scenarios. Thus, VLBI observations are the only way to break the degeneracy of the exact cause of the change in flux density in this component, assuming it is not directly related to the AGN. For this purpose, we have obtained proprietary VLBI and EVN data (PI: Panessa, project EP113) which we will publish in a subsequent paper (Panessa et al. in prep.).

\subsection{Testing the adiabatic expansion hypothesis in component C3}
Component C3 (C) is the only component in the data to have decreased in flux density between the two epochs of data, from 6.32$\pm$0.61 to 4.56$\pm$0.26\,mJy/beam, though we note that this only represents a change of 2.7$\sigma$. \citet{Williams4151} attribute the decrease in flux density to adiabatic expansion of component C3, as the timescales of simple radiative losses from synchrotron emission were too long to explain the flux density decrease. On the assumption that the decrease in flux density is real, we perform the same analysis here, first working out the equipartition magnetic field, and subsequently the synchrotron cooling timescale and compare that with the predicted flux density decrease from adiabatic expansion. A description of this method is found in Section 3.4 and 3.5 of \citet{Williams4151}. We note that while the two epochs have slightly different central frequencies, this difference would only lead to a 1 \textit{per cent} change in flux density.


Using the component size for C3 (58.1mas $\times$ 20.6mas) from the present 5\,GHz e-MERLIN observations, and following the method outlined in \citet{Williams4151}, we derive a magnetic field strength of 0.9\,mG and a synchrotron cooling timescale ($\tau \propto$ B$^{-1.5} \nu^{-0.5}$) of $\sim$2.17$\times$10$^{4}$\,years. We note, however, that a tighter limit of 700 years is still provided by the VLBI observations \citep{Ulvestad98,Mundell03,Ulvestad2005} which were discussed in \citet{Williams4151}. We therefore reaffirm that synchrotron cooling cannot explain the flux density decrease in component C3.

We now turn to understanding the flux density decrease from adiabatic expansion of component C3. Following \citet{ScheuerWilliams}, the change in flux density is related to a simple linear expansion factor, F, and the spectral index, $\alpha$, as F$^{4\alpha -2}$. Therefore, to explain the change in flux density of $\sim$26 \textit{per cent} between the two epochs, we require a linear expansion factor of F$\sim$1.04-1.05. which is a similar value to that calculated by \citet{Williams4151}, and yields an expansion velocity of $\sim$1700km s$^{-1}$, qualitatively the same as that calculated by \citet{Williams4151}.

\subsection{NGC\,4151 and jets in other low luminosity AGN}

Radio jets are seen in other low luminosity AGN, e.g. NGC\,1052 \citep{Baczko2016,Baczko2019,Nakahara2020}. However there are significant differences between both the jet properties and the host galaxy properties between NGC\,4151 and these other AGN. For example, in NGC\,1052 jet speeds of $\sim$0.34c (western jet) and $\sim$0.53c (eastern jet) are found whereas in NGC\,4151 no detectable movement is found \citep{Ulvestad2005}. Also in NGC\,1052, the overall morphologies change over a four year period and flux density variations in jet components are seen on that timescale \citep{Baczko2016,Baczko2019}. Moreover the radio luminosity of NGC\,1052 is 100 $\times$ that of NGC\,4151 and is unequivocally a radio-loud object \citep{Kadler2004}. However although NGC\,4151 and NGC\,1052 have similar black hole masses ($\rm log (M_{BH})$ = 7.7 and 8.2, respectively), they reside in very different host galaxies: NGC\,4151 is a Seyfert 1.5 spiral galaxy whereas NGC\,1052 is classified as an elliptical LINER galaxy. Thus the majority of the nuclear emission in NGC\,4151 is probably powered by an efficient accretion flow whereas that in NGC\,1052 is likely powered by an inefficient ADAF flow which more naturally launches jets \citep[e.g. see][and references therein]{NarayanYi,YuanNarayan2014}. 
For a Seyfert galaxy, NGC\,4151 is particularly radio luminous \citep{nagar05}. Although usually classed as `radio-quiet', it is occasionally classed as radio loud  \citep{4151BrightestRadioQuietAGN,Kadler2004}. It also has a relatively low accretion rate ($\sim$2 \textit{per cent} Eddington \citep{McHardy4593} for a Seyfert galaxy and so may be a transition object between radio loud LINERs and radio quiet Seyferts \citep{Mahmoud2020}.

\subsection{Resolving new components along the jet}

The higher sensitivity of our new 5\,GHz e-MERLIN images have allowed us to clearly resolve compact radio components along the jet, such as C1, C2, C5 and C6, seen previously in the lower-resolution studies. The images presented in \citealt{Pedlar93} are more sensitive to the diffuse emission at 5\,GHz, but, due to the higher frequency and lack of \textit{uv}-spacings below 400k$\lambda$ in our new e-MERLIN images, the diffuse emission is resolved out. It is likely that the jet is continuous, as seen in the 21\,cm VLBA and phased VLA images of \citealt{Mundell03}, which are sensitive to a similar angular scale (40mas synthesized beam) as our data presented here along the jet spine in Fig.~\ref{fig:BigMap}. Hence, our images show compact hot spots along the jet spine, which are embedded in much larger diffuse components which are resolved out in our images.


As the 21\,cm \citet{Mundell03} data are of a similar resolution to our 5\,GHz e-MERLIN data, we are able to reach general conclusions about the individual components along the jet axis, by considering the radio spectra between the two datasets. All of the components in Fig.~\ref{fig:BigMap} appear to align closely with components in the \citet{Mundell03} image and have flux densities in the range $\sim$1-2\,mJy/beam. Given that the peak flux densities in all of these components are $1\,$\,mJy\,beam$^{-1}$ in the 5\,GHz e-MERLIN data, this implies a steeper spectral index ($\alpha$ < 0). For example, for the brightest and faintest e-MERLIN components in either the Western or Eastern jets, C5E and C2W respectively, the spectral index ranges from $\sim$-0.3 to $\sim$-1.0. These values are only representative, but are qualitatively consistent with those found by \citet{Williams4151} for the same components. The components along the jet spine are likely impact regions where the jet interacts with the extended emission line region and their steeper spectral indices are likely indicative of synchrotron ageing due to radiative losses \citep{Condon2016}.

\section{Conclusions}

We present new 5\,GHz e-MERLIN observations of the radio-quiet Seyfert galaxy NGC\,4151, achieving 70$\mu$Jy rms-sensitivity and $\sim$0.05$\arcsec$ resolution along the jet axis. We find that the Eastern component, C4E, is responsible for the majority of the increase in flux density in the nucleus of NGC\,4151 reported in \citet{Williams4151}. Component C4W, generally favoured as the most likely location of the AGN, has not varied significantly between epochs (2.6$\sigma$ increase). We compare the flux density to archival MERLIN data from 1991 \citep{Pedlar93} and find that when comparing like-for-like \textit{uv}-coverage, component C4E has increased by a factor of $\sim$2, representing a 8.2$\sigma$ increase. Furthermore, we calculate in-band spectral indices for all of the detected e-MERLIN components at 5\,GHz, which show that component C4W is consistent with being spectrally flat ($\alpha$=-0.02$\pm$0.34), whereas component C4E is slightly steeper ($\alpha$=-0.19$\pm$0.17) and C3 is spectrally steeper still ($\alpha$=-1.19$\pm$0.76). Owing to the improved sensitivity of e-MERLIN, component X is detected in the nuclear region for the first time at 5\,GHz, but due to the low flux density, we are unable to produce a reliable spectral index image or variability constraints. 

We interpret the increase in flux density of component C4E as most likely due to a region of particle acceleration or continued jet-ELR interaction causing shock excited radio emission. This scenario will be tested using further high-resolution radio interferometers such as the VLBA and EVN. We emphasise that multiple frequency observations are best for future radio studies, as the radio spectral index, $\alpha$, will be important in discerning the cause of the increase in radio emission. We also test the adiabatic expansion hypothesis suggested by \citet{Williams4151}, and find it consistent with our new data, though we note that the flux decrease of this component is only significant to 2.7$\sigma$. Furthermore, to probe the compact regions further along the jet spine than previously performed at this frequency with the MERLIN interferometer. By comparing to the VLBA/VLA data of \citet{Mundell03}, with a similar angular resolution as e-MERLIN at 5\,GHz, we are able to give rough estimates on the spectral index for these components, which indicate that these are likely hotspots of interaction regions along the jet spine.
These hotspots are likely embedded within a more diffuse structure as the jet of NGC\,4151 is continuous as shown by \citet{Mundell03} and other studies with higher sensitivity from longer observations. The inclusion of the Lovell and possibly the VLA would be needed to explore the diffuse emission futher with e-MERLIN. 

This work emphasises the need for high-resolution studies to fully understand the nature of variability in LLAGN. Previous radio variability studies of LLAGN have had mixed results \citep{wrobel00,Nagar2002,anderson05,Mundell2009,SadieJones2011,Bell2011,Baldi2015}, providing evidence for and against nuclear variability. For the most part, these studies have assumed that the variability is due only to the AGN. However, our observations of NGC\,4151 show the need for high-resolution radio interferometry in nearby LLAGN as it is plausible that previous studies of changes in flux density are are potentially unrelated to AGN activity, and can be attributed to regions in the jet. Such variability caused by indirect AGN activity can have important implications for waveband scaling relationships for AGN and therefore further investigation is needed to fully understand the origin of variability in radio-quiet AGN.

\section*{Acknowledgements}

We thank the anonymous reviewer for their comments and revisions, which greatly improved the quality of this manuscript. 
We acknowledge funding from the Mayflower Scholarship from the University of Southampton afforded to DW to complete this work. This work was supported by the Oxford Centre for Astrophysical
Surveys, which is funded through generous support from the Hintze
Family Charitable Foundation. The research leading to these results has received funding from the European Union’s Horizon 2020 Programme under the AHEAD project (grant agreement n. 654215). This publication has also received funding from the European Union's Horizon 2020 research and innovation programme under grant agreement No 730562 [RadioNet]. IMcH thanks the Royal Society for the award of a Royal Society Leverhulme Trust Senior Research Fellowship. RDB and IMcH also acknowledge the support of STFC under grant [ST/M001326/1]. FP acknowledges support from grant PRIN-INAF SKA-CTA 2016. GB acknowledges financial support under the INTEGRAL ASI-INAF agreement 2013-025-R1. JHK acknowledges financial support from the European Union's Horizon 2020 research and innovation programme under Marie Sk\l{}odowska-Curie grant agreement No 721463 to the SUNDIAL ITN network, and from the Spanish Ministry of Science, Innovation and Universities (MCIU) and the European Regional Development Fund (FEDER) under the grant with reference AYA2016-76219-P, from IAC project P/300724, financed by the Ministry of Science, Innovation and Universities, through the State Budget and by the Canary Islands Department of Economy, Knowledge and Employment, through the Regional Budget of the Autonomous Community, and from the Fundaci\'on BBVA under its 2017 programme of assistance to scientific research groups, for the project "Using machine-learning techniques to drag galaxies from the noise in deep imaging". DMF wishes to acknowledge funding from an STFC Q10 consolidated grant [ST/M001334/1]. EB and JW acknowledge support from the UK's Science and Technology Facilities Council [grant number ST/M503514/1] and [grant number ST/M001008/1], respectively. JM acknowledges financial support from the State Agency for Research of the Spanish MCIU through the ``Center of Excellence Severo Ochoa'' award to the Instituto de Astrof\'isica de Andaluc\'ia (SEV-2017-0709) and from the grant  RTI2018-096228-B-C31 (MICIU/FEDER, EU). CGM acknowledges financial support from STFC.  We also acknowledge Jodrell Bank Centre for Astrophysics, which is funded by the STFC. eMERLIN and formerly, MERLIN, is a National Facility operated by the University of Manchester at Jodrell Bank Observatory on behalf of STFC. MP acknowledge the support from the Royal Society Newton International Fellowship.




\bibliographystyle{mnras}
\bibliography{bib}


\bsp	
\label{lastpage}
\end{document}